\documentclass{Interspeech2024}

\usepackage{xcolor}
\usepackage{enumitem}
\usepackage{graphicx}
\definecolor{nicergreen}{rgb}{0.13, 0.54, 0.13}
\definecolor{nicered}{rgb}{0.83, 0.16, 0.16}
\definecolor{myhighlight}{rgb}{0.91, 0.95, 0.93}

\usepackage{lipsum}

\newcommand\blfootnote[1]{%
  \begingroup
  \renewcommand\thefootnote{}\footnote{#1}%
  \addtocounter{footnote}{-1}%
  \endgroup
}

\interspeechcameraready

\title{ElasticAST: An Audio Spectrogram Transformer \\ 
for All Length and Resolutions}

\name[affiliation=]{Jiu}{Feng}
\name[affiliation=]{Mehmet Hamza}{Erol}
\name[affiliation=]{Joon Son}{Chung}
\name[affiliation=]{Arda}{Senocak}

\address{
  Korea Advanced Institute of Science and Technology, South Korea}
\email{\{jiufeng2000,mehamerol,arda.senocak\}@gmail.com, joonsc@kaist.ac.kr}


\begin{document}

\maketitle

\begin{abstract}
Transformers have rapidly overtaken CNN-based architectures as the new standard in audio classification. Transformer-based models, such as the Audio Spectrogram Transformers (AST), also inherit the fixed-size input paradigm from CNNs. However, this leads to performance degradation for ASTs in the inference when input lengths vary from the training. This paper introduces an approach that enables the use of variable-length audio inputs with AST models during both training and inference. By employing sequence packing, our method ElasticAST, accommodates any audio length during training, thereby offering flexibility across all lengths and resolutions at the inference. This flexibility allows ElasticAST to maintain evaluation capabilities at various lengths or resolutions and achieve similar performance to standard ASTs trained at specific lengths or resolutions. Moreover, experiments demonstrate ElasticAST's better performance when trained and evaluated on native-length audio datasets. 
Code is available at: \href{https://github.com/JiuFengSC/ElasticAST}{\underline{https://github.com/JiuFengSC/ElasticAST}}
\blfootnote{This work was supported by the National Research Foundation of Korea (NRF) grant funded by the Korea government (MSIT) (No. RS-2023-00212845).
}
\end{abstract}

\vspace{-2mm}\section{Introduction}
\label{sec:intro}
Until not long ago, convolution-based neural networks were the prominent approach in both computer vision \cite{he2016deep,yolo} and audio processing \cite{kong2020panns}. More recently, transformers \cite{vaswani2017attention} have made a significant impact on numerous computer vision \cite{Dosovitskiy2021vit,Girdhar2019video,lu2019vilbert,Carion2020EndtoEndOD,caron2021dino,touvron2021training,zheng2021rethinking} and audio processing tasks \cite{gong21b_interspeech,gong2022ssast,koutini2021efficient,baade2022mae,huangmasked,chong2022masked,nagrani2021attention,hsu2021hubert,Baevski2020wav2vec2A}, and CNN-based architectures are being replaced with these attention-based architectures. 
Despite this replacement, transformers still adhere to the fixed-size input paradigm as CNNs do. Similarly, Audio Spectrogram Transformers (ASTs)~\cite{gong21b_interspeech} take a fixed-size input where the input spectrograms are divided into fixed-size patches to create tokens as input for the transformer encoder. To obtain a fixed-size input, audio spectrograms are either trimmed or padded to a fixed size. Considering transformers can process any sequence length, using varying input sizes rather than fixed ones, can be a more optimal and natural choice for audio processing tasks. 

This fixed input size paradigm poses several challenges and flaws: (1) Recent datasets, such as VoxCeleb and Epic-Sounds, consist of audio recordings of various lengths (see Figure~\ref{fig:motivation}). Considering the recent developments in using in-the-wild data, self-supervised learning, and multimodal learning, having data in various lengths becomes quite natural. (2) Trimming or padding the input data is a suboptimal choice, as it can easily lead to the discarding or contamination of information. (3) AST-based models lack flexibility when evaluated with inputs of different lengths or temporal resolutions, both of which result in varying sequence lengths, compared to those used during training. This necessitates training different AST models for circumstances with varying sequence length requirements, such as computation budget or memory consumption. Overall, these considerations make standard ASTs relatively limited. Therefore, it is appealing to explore the flexibility of sequence lengths in Audio Spectrogram Transformers. Specifically, this involves designing a single AST model capable of handling variable input sequence lengths during both training and inference.

In this work, we present ElasticAST, an Audio Spectrogram Transformer model that turns standard AST into a model capable of processing audio of any length or temporal resolution during both training and inference without trimming or padding. The resulting model is functionally superior to standard ASTs as it maintains its performance across various lengths or resolutions of audio during the inference stage and also delivers better performance on datasets containing audio of various lengths. 

\begin{figure}[t!]
\centering
{
\resizebox{0.9\linewidth}{!}{%
\begin{tabular}{c}
\includegraphics[width = 1.0\linewidth]{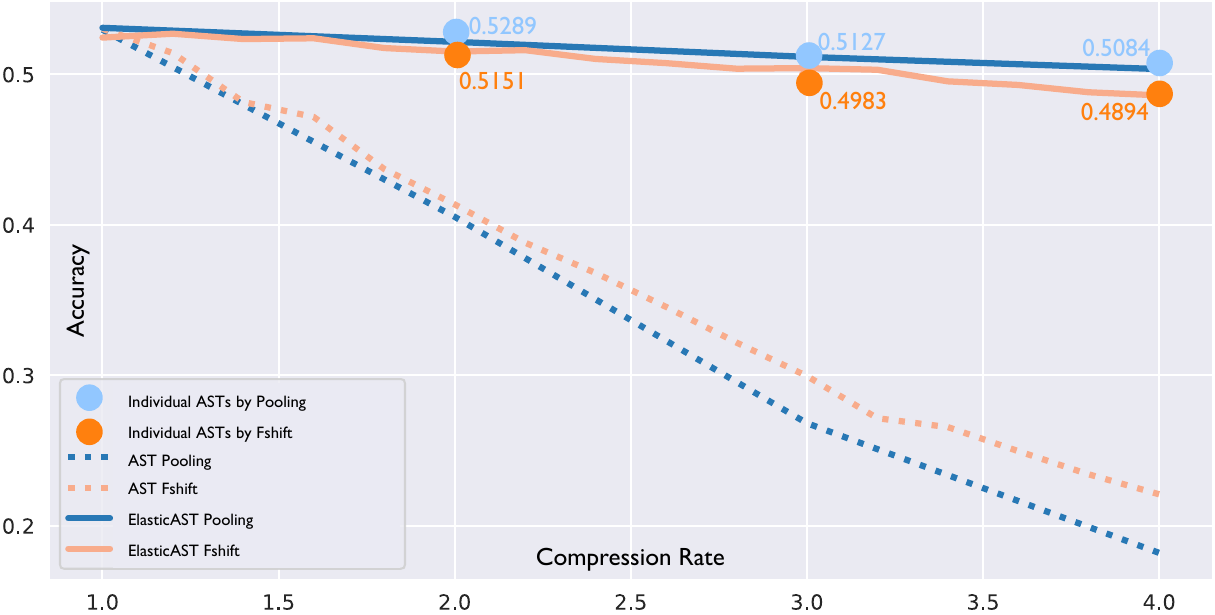} \\ 
\end{tabular}
}
}
\vspace{-4mm}
\caption{\textbf{Standard ASTs vs. ElasticAST.} Standard ASTs' performance degrades when evaluated on audio lengths different from their trained lengths, while ElasticAST remains flexible to varying lengths.}
\label{fig:teaser}
\vspace{-6mm}
\end{figure}

\begin{figure}[t!
]
\centering
{
\resizebox{\linewidth}{!}{%
\begin{tabular}{c}
\includegraphics[width = 1.0\linewidth]{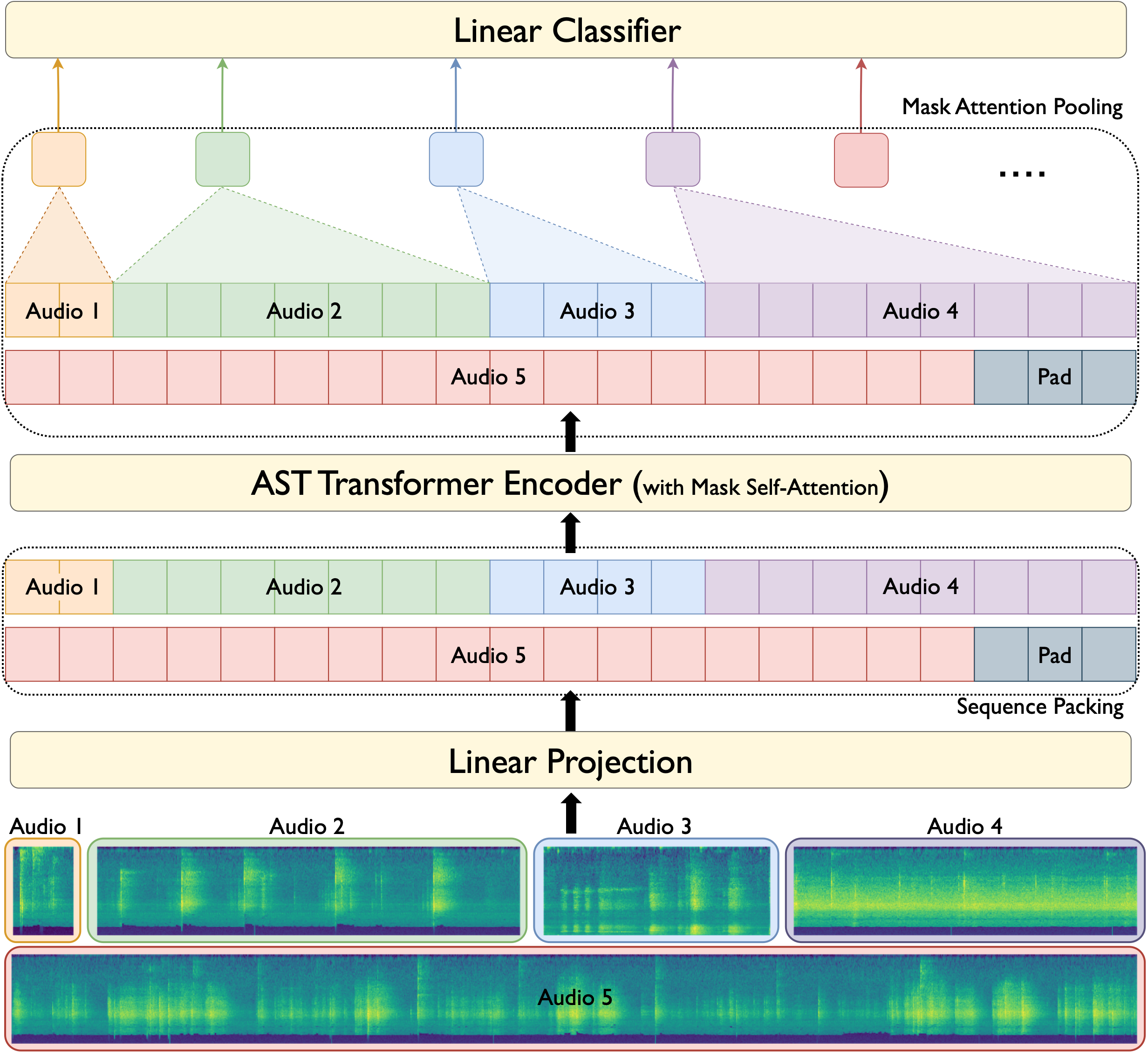} \\
\end{tabular}
}
}
\vspace{-4mm}
\caption{\textbf{Our ElasticAST framework.}}
\label{fig:pipeline}
\vspace{-6mm}
\end{figure}

There are previous approaches that explore different sequence lengths or temporal resolutions in AST-based architectures.~\cite{feng2023flexiast,beyer2023flexivit} focus on providing patch-size flexibility to ASTs and ViTs. Different patch sizes lead to different sequence lengths. Patch sizes are randomly selected during training, and a resizing algorithm is applied to convert the patch embedding weights accordingly for the different patch sizes. As a result, models gain flexibility with different patch sizes during the inference stage. Our method follows a similar direction in terms of providing flexibility to sequence length during both training and inference stages. However, instead of patch sizes, our model focuses on variable lengths and temporal resolutions of audio inputs. Another related work is~\cite{feng2024coarse}, which uses mixed resolutions of audios to train ASTs efficiently. The main idea is to process lower-resolution audio (coarse) early in training, and then fine-tune with high-resolution data (fine) later in a hierarchical manner. Rather than offering flexibility, this work focuses on training efficiency. In contrast, our model randomly mixes various resolutions of audios during training at once without employing any curriculum learning strategy, making the model capable of seamlessly processing any temporal resolution of audio in the inference. Moreover, both of these previous works still use fixed length input.
The work most similar to ours is NaViT~\cite{dehghani2024patch}, which allows Vision Transformers to handle a variety of image resolutions. Inspired by the discoveries in this study, we investigate appropriate methods to make adaptations to the standard ASTs for flexibility across all lengths and temporal resolutions of audio. To accomplish this task, we take the following steps (shown in Figure~\ref{fig:pipeline}): (1) Instead of using a fixed audio length or resolution during training, we randomly select the resolution or use the variable native length of the audios without trimming or padding. (2) We pack these audios of different lengths (or resolutions) into a single sequence to process them all at once during training. (3) We use the standard AST architecture with minimal changes, such as limiting the scope of attention to each individual sample to prevent it from attending to other samples within a packed sequence, and replacing the class tokens with a masked attention pooling mechanism. The rest of the model remains identical. We summarize the contributions of our work as follows:

\begin{figure}[t!]
\centering
{
\resizebox{0.9\linewidth}{!}{%
\begin{tabular}{c}
\includegraphics[width = 1.0\linewidth]{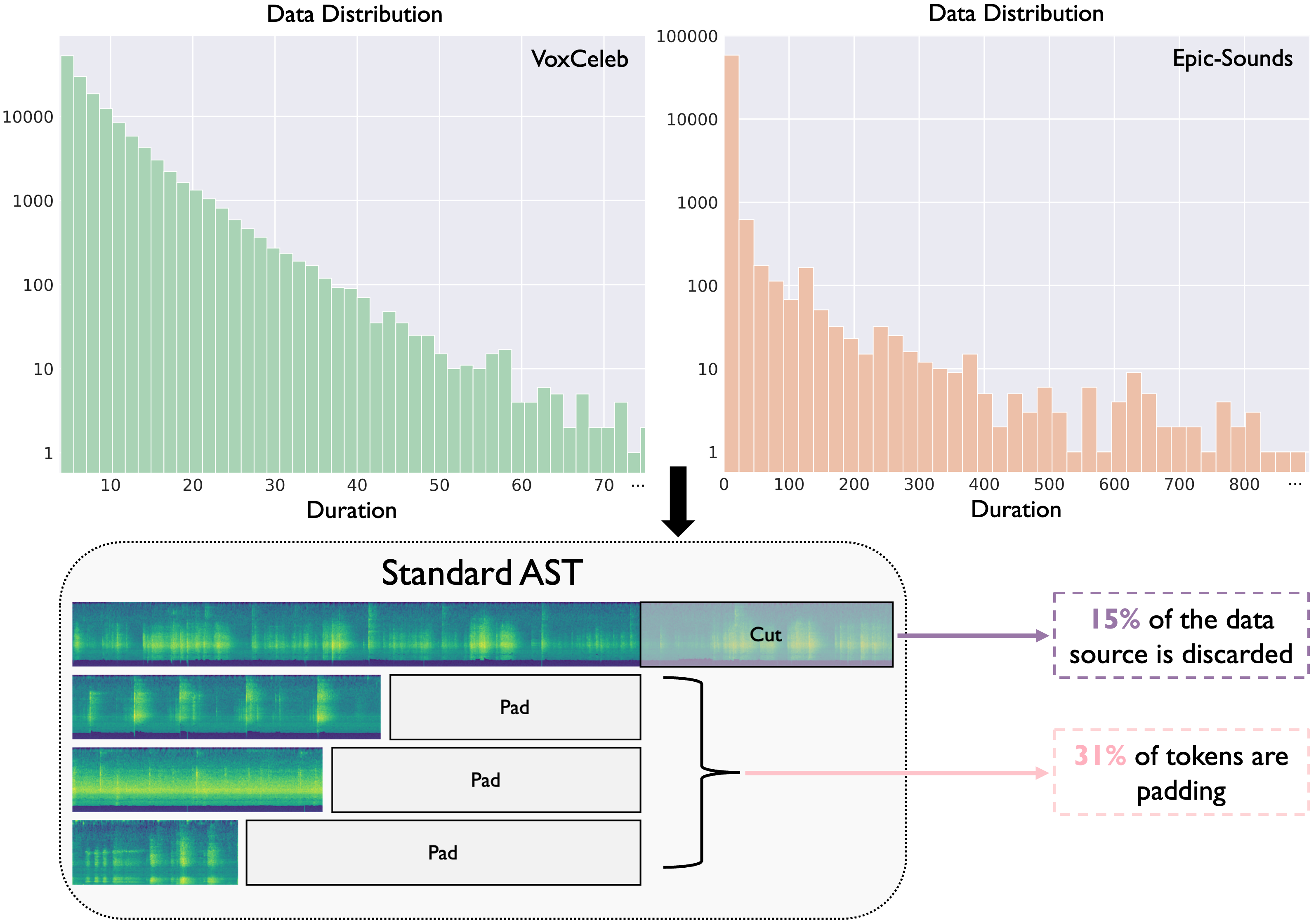} \\
\end{tabular}
}
}
\vspace{-4mm}
\caption{\textbf{Variable Length Datasets and standard AST training input.} Due to the fixed-length processing constraints, ASTs discard informative tokens and introduce non-informative tokens.}
\label{fig:motivation}
\vspace{-6mm}
\end{figure}

\begin{itemize}[noitemsep, topsep=0pt]
\item We demonstrate that standard Audio Spectrogram Transformers (ASTs) lack the flexibility to be trained and evaluated on variable lengths or resolutions different from those on which they were initially trained.
\item We introduce an approach that enables the creation of ElasticAST which allows standard ASTs to be trainable with variable native lengths or temporal resolutions of audio.
\item ElasticAST is a single AST model capable of operating across all lengths and resolutions at the inference stage without significant performance degradation, while achieving performance comparable to standard ASTs trained at fixed lengths or resolutions.
\item We show that, in addition to flexible model usage, ElasticAST can surpass the performance of standard ASTs on datasets of variable lengths, such as VoxCeleb and Epic-Sounds, by leveraging the full semantic content of audio without the need for cutting or padding.

\end{itemize}

\vspace{-2mm}\section{Approach}
\label{sec:approach}

\vspace{-2mm}\subsection{Preliminaries}
\vspace{-1mm}
The Audio Spectrogram Transformer (AST \cite{gong21b_interspeech}) utilizes a transformer-based architecture to process audio spectrograms. It starts by splitting each spectrogram $x \in \mathbb{R}^{f \times t}$ from a batch $X \in \mathbb{R}^{B \times f \times t}$ into a sequence of $S$ smaller patches: $x \to x_i \in \mathbb{R}^{p \times p}$, where $i$ ranges from $0$ to $S$ and $S=(f/p)\times (t/p)$. These patches are then converted into embeddings via a linear projection layer $e_i = P(x_{i}) \in \mathbb{R}^D$. Next, a special {\fontfamily{pcr}\selectfont [cls]} token is prepended to the sequence, increasing the length to $N=S+1$. The learnable positional embeddings are added to provide the order information of the tokens. The sequences are then transformed by the transformer encoder. The encoder's output for the {\fontfamily{pcr}\selectfont [cls]} token acts as the audio spectrogram's representation for downstream tasks, such as classification.
\vspace{-2mm}\subsection{Architectural changes}
\vspace{-2mm}
ElasticAST is a conceptually simple extension of standard ASTs, designed to accommodate the use of various audio spectrogram lengths during both the training and inference stages. This flexibility is achieved with minimal architectural modifications to conventional ASTs. 

\noindent\textbf{Sequence Packing.} 
Unlike AST, which allocates $N$ tokens for all the $B$ input spectrograms, our model employs a sequence packing method that accommodates the varying lengths by organizing the spectrograms into token sequences $X \in \mathbb{R}^{B' \times N' \times D}$ after patchification as illustrated in Figure \ref{fig:pipeline}.
This method initiates by sequentially filling the first token sequence row with tokens derived from the samples until reaching the preset token limit per row, $L'$ (as a default set to 2048). If the tokens from a sample about to surpass this limit, they are deferred to the subsequent row, and the allocation process continues accordingly for the remaining samples.
This packing technique generates $B'$ rows of token sequences, each potentially varying in length but constrained by $L'$. To facilitate processing by the transformer encoder, as shown in Figure \ref{fig:pipeline}, we introduce minimal padding tokens to each row, standardizing their lengths to $N' \leq L'$, which corresponds to the longest sequence length among the rows.
Our approach to packing is deliberately straightforward, prioritizing simplicity by processing samples sequentially in their original order and filling rows on a first-come, first-served basis. Future work may investigate more sophisticated packing algorithms to enhance efficiency.

\noindent\textbf{Mask Self-Attention.}
In conventional patch-based transformer models \cite{gong21b_interspeech,Dosovitskiy2021vit}, only the tokens from the same sample form a row, and these tokens are aware of each other within the transformer encoder through global attention, formalized as $\text{Attention}(Q, K, V) = \text{softmax}\left(QK^T\right) \cdot V$ where $Q, K, V \in \mathbb{R}^{B \times N \times D}$.
However, this mechanism becomes impractical when sequences are comprised of tokens from different samples, as it fails to constrain the attention within the tokens of the same sample.
To address this, our architecture is designed to prevent tokens from different samples within the same sequence from attending to one another. We achieve this by introducing a Masked Self-Attention mechanism. The main idea is to introduce a boolean mask $M \in \mathbb{B}^{B' \times N' \times N'}$ for each sequence in a batch, where $M_{b,i,j}$ represents if the $i$th token should attend to th $j$th token in $b$th sequence, thus encoding within sample attention. Then, we modify the original attention mechanism as: $\text{Attention}(Q, K, V) = \text{softmax}\left(\text{Mask}(QK^T)\right) \cdot V$ where $Q, K, V \in \mathbb{R}^{B' \times N' \times D}$, by selectively preventing cross-sample attention, visually represented by different colors in Figure \ref{fig:pipeline}.

\noindent\textbf{Mask Attention Pooling.}
After the sequences are processed by the transformer encoder without cross-sample attention contamination, the representation of each sample before the linear classifier is obtained through Mask Attention Pooling \cite{Zhai0HB22}.
In AST, each sequence row has a prepended {\fontfamily{pcr}\selectfont [cls]} token used as the representation of that sample. For the ElasticAST, rather than coupling a {\fontfamily{pcr}\selectfont [cls]} token with each packed sample, we employ a Mask Attention Pooling layer on the top of the encoder to derive sample representations, serving as an effective alternative.
This architecture mirrors the previously described Mask Self-Attention mechanism, wherein the $K, V \in \mathbb{R}^{B' \times N' \times D}$ matrices are obtained from the input, while another $Q' \in \mathbb{R}^{B' \times n \times D}$ matrix is generated from the repetition of a learnable query vector parameter $q \in \mathbb{R}^{D}$, where $n$ refers to the maximum number of packed samples in all rows.
Then, with a generated mask $M \in \mathbb{B}^{B' \times n \times N'}$ where $M_{b,i,j}$ corresponds to if $j$th token in the $b$th sequence belongs to the $i$th sample. Then, the mask attention pooling, $\text{AttnPool} = \text{softmax}\left(\text{Mask}(Q'K^T)\right) \cdot V$, subsequently extracts representations $X \in \mathbb{R}^{B' \times n' \times D}$, in a similar way of employing masking to ensure pooling is confined to tokens within a sample, as depicted by the cone shadow areas in Figure \ref{fig:pipeline}. Afterward, the obtained sequence is unpacked into $X \in \mathbb{R}^{B \times D}$ by reordering the representations of each sample, treating them as the {\fontfamily{pcr}\selectfont [cls]}, and fed into the linear classifier. 
\\

\vspace{-6mm}\section{Experiments}
\subsection{Datasets and Evaluation Metrics}\label{sec:dataset}
\vspace{-2mm}
\noindent\textbf{Datasets.} In our experiments, we utilize four datasets: AudioSet, VGGSound, VoxCeleb, and Epic-Sounds. AudioSet~\cite{gemmeke2017audio} is a large multi-label dataset with approximately 2 million 10-second clips spanning 527 labels across diverse audio categories. VGGSound~\cite{VGGSound} includes around 200,000 10-second video clips, labeled with 309 sound classes such as objects and human activities. VoxCeleb~\cite{nagrani2020voxceleb} is an audio-visual dataset focused on human speech, featuring 1,251 speakers and approximately 145,000 utterances across a range of durations from 4 to 144 seconds. For experimental purposes, we impose a hypothetical upper limit of 30 seconds on this dataset to simplify the experiments and manage the memory usage 
effectively. Lastly, Epic-Sounds~\cite{EPICSOUNDS2023}, derived from first-person (egocentric) videos, includes 44 categories and a total of 75.9k audio files of various lengths (see Figure~\ref{fig:motivation}). Similar to VoxCeleb, we set a provisional maximum duration of 30 seconds for this dataset, though this limit can be adjusted as needed.

\noindent\textbf{Evaluation metrics.} Given the presence of multi-labels in each AudioSet sample, we use mean average precision (mAP) for evaluation across all categories. For the other datasets, we report the Top-1 classification accuracy (Acc) as our measure of evaluation since each sample has only a single label.
\vspace{-2mm}
\subsection{Implementation Details}
\vspace{-2mm}
In this paper, the configuration for standard ASTs follows the same choices as those in~\cite{feng2023flexiast}. The batch and the patch size are set to 12 and 16 (B/16) respectively, and all models are initialized with ViT~\cite{dosovitskiyimage} (ImageNet Pretrained) for every dataset, except for VoxCeleb, where SSAST~\cite{gong2022ssast} weights are employed. The learning rate is established at 1e-5 for all datasets, apart from the Epic-Sounds dataset, which is set at 1e-4 by following~\cite{EPICSOUNDS2023}. To accommodate varying audio lengths, we shifted from a 1D to a 2D positional \cite{Pix2Struct,RonenLG23} embeddings, encoding frequency and time positions separately.
Our ElasticAST uses the identical settings as the standard ASTs in this paper. Additional minor architectural changes are already discussed in Section~\ref{sec:approach}. By default, we use a window of 25 ms with a frame shift of 10 ms to transform waveforms into 128 mel-fbank. The resulting mel-spectrogram shapes for 10-second audio clips are as follows: for AudioSet and VGGSound, 128 × 1024. For variable-length audio clips (in VoxCeleb and Epic-Sounds), the resulting temporal length dimension is different according to the native length of the audios. We adjust the frame shift (Fshift) hyperparameter for each different audio resolutions so that the temporal length dimension changes accordingly.
All the results that are used to draw graphs in this paper are available anonymously at \href{https://sites.google.com/view/elasticast-interpseech24}{https://sites.google.com/view/elasticast-interpseech24}.

\begin{figure}[t!]
\centering
{
\resizebox{\linewidth}{!}{%
\begin{tabular}{c}
\includegraphics[width = 1.0\linewidth]{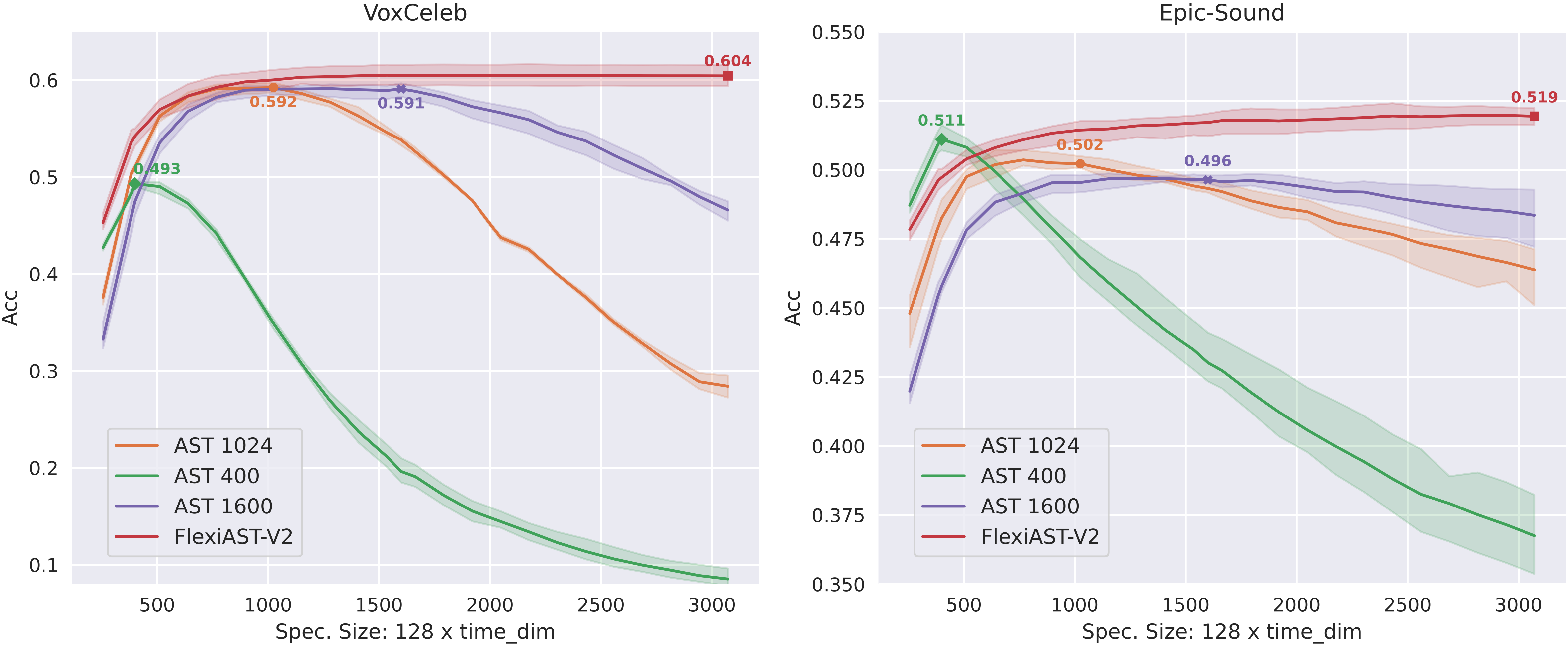} \\
\end{tabular}
}
}
\vspace{-4mm}
\caption{\textbf{Results on variable native length audios.}}
\label{fig:len_flex}
\vspace{-6mm}
\end{figure}
\vspace{-2mm}
\subsection{Results of Various Length Training}
\vspace{-1mm}
We perform experiments with various natural lengths of audio to assess the potential of ElasticAST. Thus, we use VoxCeleb and Epic-Sounds datasets, which consist of various lengths of audio as shown in Figure~\ref{fig:motivation}, for the experiments in this section. For the sake of memory, the maximum length of the audio is limited to 30 seconds (see Section~\ref{sec:dataset}). ElasticAST is trained at the native length of the audio by packing them into batches of sequences without cutting or padding. Meanwhile, AST models are trained at fixed length by trimming and padding as they can not accommodate varying input lengths during training. We compare the performance of ElasticAST to various AST models in two perspectives. First, We evaluate the performance of ElasticAST by using the native length of the audios in the inference stage without any alteration, while standard ASTs are evaluated with the audio lengths at which they are trained. All of these results are depicted with scattered dots in Figure~\ref{fig:len_flex}. 
Second, we evaluate every model at a series of audio lengths $x_i$ from 256 to 3072, such that standard ASTs always apply cutting or padding whenever the native length of the audio is not $x_i$. However, our model only cuts the audio if the native length is longer than $x_i$; otherwise, no padding is applied, and the native audio is processed as is.

\noindent\textbf{Summary.} (1) ElasticAST performs well with various data lengths. In contrast, when AST models are evaluated at a different time length than the ones used during training, their performance collapses. (2) ElasticAST on native length audio can surpass the performance of all standard ASTs by leveraging the full semantic content of audio without the need for cutting or padding (indicated by scattered dots). 
(3) During inference, with the increasing audio length, more semantics are provided, and ElasticAST can adapt to this change by utilizing a larger semantic volume to boost its performance towards saturation. The more data it can process, the better it performs. However, AST models do not necessarily improve their performance. We hypothesize that the semantic volume that AST can handle is fixed during training.
(4) During the training of ElasticAST, no information in the dataset is discarded by trimming, and meanwhile, fewer padded tokens are used, as shown in Table~\ref{tab:pool_pad}. However, the AST model has to trim longer audio or add non-informative padding tokens to fit them into its length requirement. Given the same amount of calculated tokens, ElasticAST consumes a larger meaningful semantic volume. This highlights the resource consumption efficiency of our architecture.

\vspace{-2mm}
\subsection{Results of Various Temporal Resolutions Training}\label{sec:fshift}
\begin{figure}[t!]
\centering
{
\resizebox{\linewidth}{!}{%
\begin{tabular}{c}
\includegraphics[width = 1.0\linewidth]{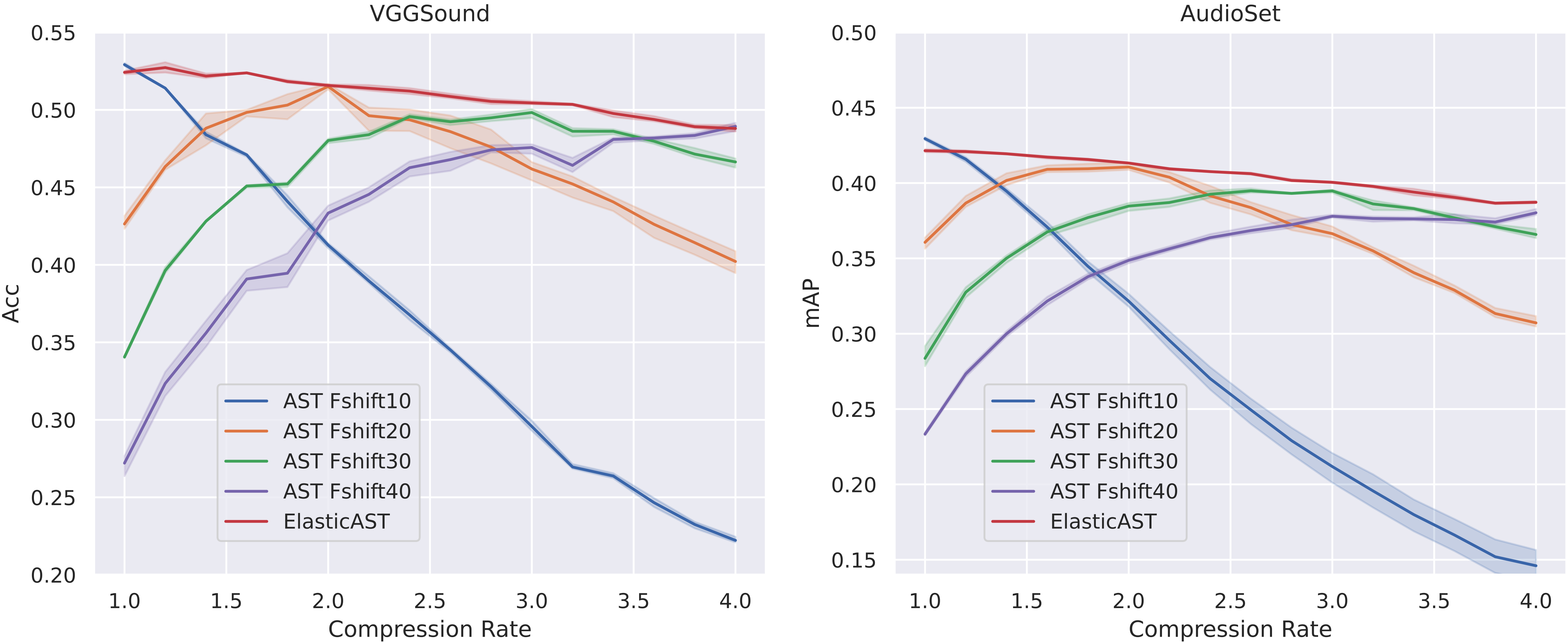} \\
\end{tabular}
}
}
\vspace{-4mm}
\caption{\textbf{Results on various resolutions.}}
\label{fig:res_flex}
\vspace{-6mm}
\end{figure}

\vspace{-1mm}
This section presents the results of ElasticAST using various temporal resolutions. We parameterize the temporal resolution through the Fshift (frame shift) value when generating spectrograms~\cite{kazakos2021slow,feng2024coarse}. This value determines the length and, consequently, the detail/resolution of the spectrogram. We temporally compress the mel-spectrograms by a factor of $C_i$, where the frame-shift value is multiplied by randomly sampled $C_i$ from $C = \{1.0,1.2,1,4,...3.8,4.0\}$ during training (mixed-resolution training). Conventional ASTs are trained with a fixed resolution since they cannot accommodate varying input lengths during training. We compare our ElasticAST to standard ASTs during inference across multiple compression rates on VGGSound and AudioSet, as illustrated in Figure~\ref{fig:res_flex}. The rationale behind choosing these two datasets is that they consist of fixed-length audios, and we convert them into various lengths through different compression rates (Fshift values).

\noindent\textbf{Summary.} (1) Our results show that ElasticAST possesses significant flexibility in handling a wide range of resolutions across both datasets. In comparison, the performance of standard ASTs declines when tested at resolutions different from those used in training. (2) Beyond flexibility, our model generally matches or surpasses the performance of standard ASTs, even on the resolutions for which standard ASTs were specifically trained. (3) Our model allows for the training of a single model adaptable to various resolutions, unlike standard ASTs, which must be trained individually for each resolution. This means that, within a given inference budget, ElasticAST can seamlessly adjust to any computational budget, whereas standard ASTs would require training a new model for each specific computational scenario.

\vspace{-2mm}
\subsection{Ablation Study}
\vspace{-2mm}

The main focus of ElasticAST is to provide flexibility to AST models for using variable time length audio input in both the training and inference stages. In Section~\ref{sec:fshift}, we present the results of using various temporal resolutions. The time length is changed through the Fshift operation. Additionally, there can be other methods for reducing temporal length~\cite{liu2022simple,feng2024coarse}. In this ablation study, we investigate \textit{average pooling} strategy in ElasticAST for using various temporal lengths. By following~\cite{feng2024coarse}, a mel-spectrogram is processed through an extra layer of average-pooling. This layer uses a kernel and stride both sized $1 \times C$, effectively reducing the mel-spectrogram's temporal dimension by a factor of $C$. The training and experimental settings are identical to those described in Section~\ref{sec:fshift}, except for the new reduction method, \textit{average pooling}, which requires integer numbers. Therefore, only the values $C = \{1, 2, 3, 4\}$ are used as compression rates. To save computational time and resource, we conduct this experiment on VGGSound. The results are shown in the right side of Table~\ref{tab:pool_pad}. Similar to the Fshift approach, ElasticAST also demonstrates flexibility in handling various range of temporal lengths with this method.
\vspace{-2mm}
\subsection{Analysis on Token Usage}
\vspace{-2mm}
In this section we analyze the token usage efficiency of ElasticAST and standard ASTs when various length audios are given during training. Due to fixed-length processing constraints, StandardASTs either cut or pad the inputs. 
In contrast, ElasticAST flexibly handles audios of varying lengths without resorting to cutting or padding operations. 
However, as described in Section 2, when packing audios of different lengths into a sequence, we use padding to standardize the length of the token sequence rows, $L'$. Using the VoxCeleb dataset, our findings are presented in Table~\ref{tab:pool_pad}. The results reveal that StandardAST introduces non-informative padding tokens, which occupy 31.1\% of the total training tokens, and cut 15.2\% of the tokens in training set. In contrast, ElasticAST minimizes padding and applies no cutting. Moreover, as batch size increases, the amount of padding tokens used for packing decreases. This analysis highlights ElasticAST's training efficiency in processing informative content.

\begin{table}[t!]
\footnotesize
\renewcommand{\arraystretch}{2}
\centering
    \resizebox{0.48\linewidth}{!}{
    \begin{tabular}{lcc}
    \hline
    \textbf{Models-Batch\#} & \textbf{Pad Ratio(\%)} & \textbf{Cut Ratio(\%)} \\
    \hline
    AST-All  & \textcolor{nicered}{31.1} & \textcolor{nicered}{15.2} \\
    ElasticAST-12 & \textcolor{nicergreen}{19.2} & \textcolor{nicergreen}{0} \\
    ElasticAST-64 & \textcolor{nicergreen}{14.4} & \textcolor{nicergreen}{0}  \\
    ElasticAST-128 & \textcolor{nicergreen}{13.9} & \textcolor{nicergreen}{0}  \\
    \hline
    \end{tabular}}
    \resizebox{0.5\linewidth}{!}{
    \begin{tabular}{lcc}
    \includegraphics[width=1\linewidth]{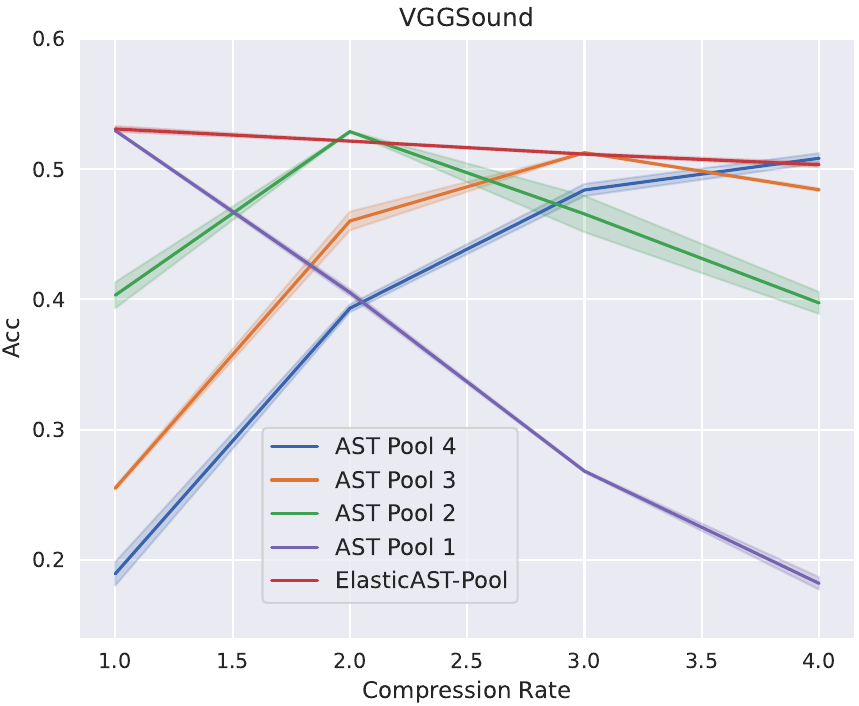}
    \end{tabular}}
        \caption{\textbf{Token Usage and Ablation Result.}}
     \vspace{-9mm}
\label{tab:pool_pad}
\end{table}

\vspace{-4mm}\section{Conclusion}
\vspace{-1mm}
In conclusion, this paper focuses on enhancing Audio Spectrogram Transformers (ASTs) to support training and inference with audios of various lengths. By introducing a strategy that employs mixed-length training and requires only minimal architectural adjustments to the AST framework, we develop \textit{ElasticAST}. This model is capable of being trained with audios of varying lengths and demonstrates flexibility without performance loss across different lengths, offering \textit{one single model for all audio lengths and resolutions}. ElasticAST's significance lies in its efficiency and adaptability to different computational budgets without the need for re-training. The ability to handle audio of various lengths is becoming increasingly important, considering the recent developments in self-supervised multimodal learning that utilizes in-the-wild data. ElasticAST's flexibility is particularly valuable for tasks involving alignment, retrieval, and generation in multimodal contexts.

\newpage
\bibliographystyle{IEEEtran}
\bibliography{mybib}

\end{document}